\documentclass[iop]{emulateapj}
\usepackage{longtable,natbib}
\usepackage[bookmarks=false]{hyperref}

\def\micron{{\mbox{$\mu{\rm m}$}}}
\def\arcsec{{\mbox{$^{\prime \prime}$}}}

\def\C{{\sl Chandra}}

\def\X{{\sl XMM \text{-} Newton}}

\def\o3{[O\small{III}]}

\begin{document}
\title{$Chandra$ Discovery of a Binary AGN in Mrk 739}

\author{Michael Koss\altaffilmark{1,2,3}, Richard Mushotzky\altaffilmark{1}, Ezequiel Treister\altaffilmark{3,4,6}, Sylvain Veilleux\altaffilmark{1}, Ranjan Vasudevan\altaffilmark{1}, Neal Miller\altaffilmark{1}, D.~B. Sanders\altaffilmark{3}, Kevin Schawinski\altaffilmark{5,6}, and Margaret Trippe\altaffilmark{1}}  
\email{koss@ifa.hawaii.edu}
\altaffiltext{1}{Astronomy Department, University of Maryland, College Park, MD, USA}
\altaffiltext{2}{Astrophysics Science Division, NASA Goddard Space Flight Center, Greenbelt, MD, USA}
\altaffiltext{3}{Institute for Astronomy, University of Hawaii, Honolulu, HI, USA}
\altaffiltext{4}{Departamento de Astronom\'{\i}a, Universidad de Concepci\'{o}n, Concepci\'{o}n, Chile}
\altaffiltext{5}{Department of Physics, Yale University, New Haven, CT, USA}
\altaffiltext{6}{Einstein Fellow}

\begin{abstract}
We have discovered  a binary AGN in the galaxy Mrk 739 using $\C$ and $Swift$ BAT.  We find two luminous ($L_{2-10 \: \mathrm{keV}}$=1.1$\times10^{43}$ and $1.0\times10^{42}$  erg s$^{-1}$), unresolved nuclei with a projected separation of 3.4 kpc (5.8$\pm$0.1$\arcsec$) coincident with two bulge components in the optical image.  The western X-ray source (Mrk 739W) is highly variable ($\times2.5$) during the 4-hour $\C$ observation and has a very hard spectrum consistent with an AGN.  While the eastern component was already known to be an AGN based on the presence of broad optical recombination lines, Mrk 739W shows no evidence of being an AGN in optical, UV, and radio observations, suggesting the critical importance of high spatial resolution hard X-ray observations ($>$$2$ keV) in finding these binary AGN.  A high level of star formation combined with a very low $L_{\mathrm{[O III]}}/L_{2-10 \: \mathrm{keV}}$ ratio cause the AGN to be missed in optical observations.  $^{12}$CO observations  of the (3--2) and (2--1) lines indicate large amounts of molecular gas in the system that could be driven towards the black holes during the violent galaxy collision and be key to fueling the binary AGN.  Mrk 739E has a high Eddington ratio of 0.71 and a small black hole (log $M_{\mathrm{BH}}$$=$$7.05\pm$0.3) consistent with an efficiently accreting AGN.  Other than NGC 6240, this stands as the nearest case of a binary AGN discovered to date.

\end{abstract}


\section{Introduction}

	The detection and frequency of binary AGN provide constraints on models of galaxy formation and an important test of the merger-driven AGN model.  Hierarchical merger models of galaxy formation and the fact that almost all massive galaxies have supermassive black holes suggest that binary black holes should be common in galaxies \citep{Volonteri:2003p11499}.  If galaxy mergers are the prime way to 'ignite' the central source by sending a large amount of gas into the center region and triggering the AGN \citep{DiMatteo:2005p5934}, then we expect some fraction of these binary black holes to be actively growing simultaneously, thus creating a binary AGN.
%possible remove galaxy formation

	Despite their theoretical importance, only a handful of close binary AGN ($<$5 kpc projected separation) have been discovered.  The two clearest cases are the \C-detected double nucleus in the luminous infrared galaxy (LIRG; $L_{\mathrm{IR}}$$>$$10^{11}$$L_{\sun}$)  NGC 6240 \citep{Komossa:2003p11219} with a projected separation of 1 kpc at a distance of 103 Mpc and the LIRG Mrk 463 at 3.8 kpc separation and a distance of 220 Mpc.  Recently some likely binary AGN have been discovered based on double-peaked [O III] $\lambda$5007 emission lines \citep{Liu:2010p7984}.  However, there is still some question whether these systems are 'true' binary AGN or are a single AGN with an asymmetric distribution of outflowing gas in the narrow line region \citep{Smith:2010p7983,Fischer:2011p11696}.  Unfortunately, these systems are at higher redshifts with extremely close separations where the resolution of $\C$ is unable to resolve these objects to confirm their binary AGN nature.

	%There has been some controversy as to whether moderate luminosity AGN are triggered through mergers, and a high frequency of dual AGN in merging galaxies supports the merger-driven AGN model.  Nearby ($z$$<$0.1) AGN selected from the SDSS have shown no merger link whereas nearby ultra hard X-ray detected AGN show a merger rate of 24\% compared to only 1\% in normal galaxies \citep{Li:2006p5063,Koss:2010p7366}. 
	
	As part of our $\C$ program of following up the close mergers detected by $Swift$ BAT  \citep{Koss:2010p7366}, we have discovered a binary AGN in Mrk 739 with a 3.4 kpc separation at a distance of 130 Mpc.     The binary AGN is particularly interesting because it shows no evidence of being an AGN in the optical, UV, or radio.  Other than NGC 6240, this stands as the nearest case of a binary AGN discovered to date.   %Section \S2 describes our imaging and spectroscopic data and the analysis techniques, \S3 describes our results and our evidence for a binary AGN, and the discussion is in \S4.
	
% A previous study of Mrk 739 using only optical emission line diagnostics found only a single AGN and a starburst nuclei (Netzer et al.~1987).   

\section{Observations and Data Analysis}
	  In the following subsections, we describe the observations and analysis of Mrk 739.  Throughout this work, we adopt $\Omega_m$= 0.3, $\Omega_\Lambda$= 0.7, and $H_0$ = 70 km s$^{-1}$ Mpc$^{-1}$ to determine distances.   At the redshift of Mrk 739, 1$\arcsec$ corresponds to 580 pc.

\subsection{Optical: SDSS Imaging and Gemini Optical Spectroscopy}
	Mrk 739 was imaged by the SDSS on March 10, 2005.  Using a S\'{e}rsic profile with a fixed bulge ($n$=4), we fit the optical nuclei using two-dimensional surface brightness fitting \citep[GALFIT;][]{Peng:2002p5550}.  In Mrk 739E, a point source component was used to measure the AGN light since it has a broad line region (BLR).
	
	We observed Mrk 739 with Gemini on February 7, 2011.  Both nuclei were observed simultaneously in the B600-G5307 grating with a 1$\arcsec$ slit in the 4300--7300 $\mathrm{\AA}$ wavelength range. The exposure totaled 37 minutes. We follow \citet{Winter:2010p6825} for correcting Milky Way reddening, starlight continuum subtraction, and fitting AGN diagnostic lines.   To correct our line ratios for extinction, we use the narrow Balmer line ratio (H$\alpha$/H$\beta$) assuming an intrinsic ratio of 3.1 and the \citet{Cardelli:1989p1821} reddening curve.
	
%  We used the standard Gemini pipeline for initial reductions and calibrations using a standard star.   

\subsection{UV and Radio: $\X$ and VLA}
Mrk 739 was observed in the UV with $\X$ in June 2009 (PI Brandt, Vasudevan in prep.).  We follow the XMM ABC guide for photometry.   We also analyzed archival VLA observations with times of 33 minutes at 1.49 GHz and  38 minutes at 4.86 GHz.

% \footnote{http://heasarc.nasa.gov/docs/xmm/abc/} 

\subsection{X-rays: $\C$}
$\C$ observed Mrk 739 on April 22, 2011 with an exposure time totaling 13 ks.   Sub pixel event repositioning was applied to improve the resolution of the image.  Two extraction regions of 1.5$\arcsec$ radius were used for spectral fitting and timing analysis with CIAO version 4.3.  To fit the X-ray spectra, we used a fixed Galactic photoelectric absorption \citep{Kalberla:2005p11562}, a floating photoelectric absorption component at $z$=$0.0297$, and a power law.  For the eastern source (Mrk 739E), we also include a pileup model because mild pileup (10\%--20\%) is expected based on pixel count rates.

\subsection{Submillimeter: CO Observations}
The $^{12}$CO (2--1) and (3--2) molecular lines of Mrk 739 were observed with the James Clerk Maxwell Telescope (JCMT) on March 12--13, 2011.  The A3 (211-279 GHz) and HARP (325-375 GHz) receivers were used.     The spectra were co-added, binned, and fitted with linear baselines.  To calculate velocity-integrated line flux densities, we assumed  an aperture efficiency of 0.60 and 0.53 for the A3 and HARP receiver and followed \citet{Greve:2009p11571}.    
% Nice paper http://www.tuc.nrao.edu/meeting/posters/benford_ele.ps
%The pointing was checked throughout each run and was  (3$\arcsec$) or better.  During the observation, spectral line standards were compared to mean JCMT values and were in excellent agreement. 
% Because the angular separation of the double nuclei is only 5.8$\arcsec$ and beam size is 15$\arcsec$ for J= 3--2  and 20.4$\arcsec$ at J= 2--1, the observation likely includes CO emission from the whole galaxy.

\section{Results}
\subsection{Optical Properties of Mrk 739}
	
\begin{figure} 
\centering
\includegraphics[width=7.7cm]{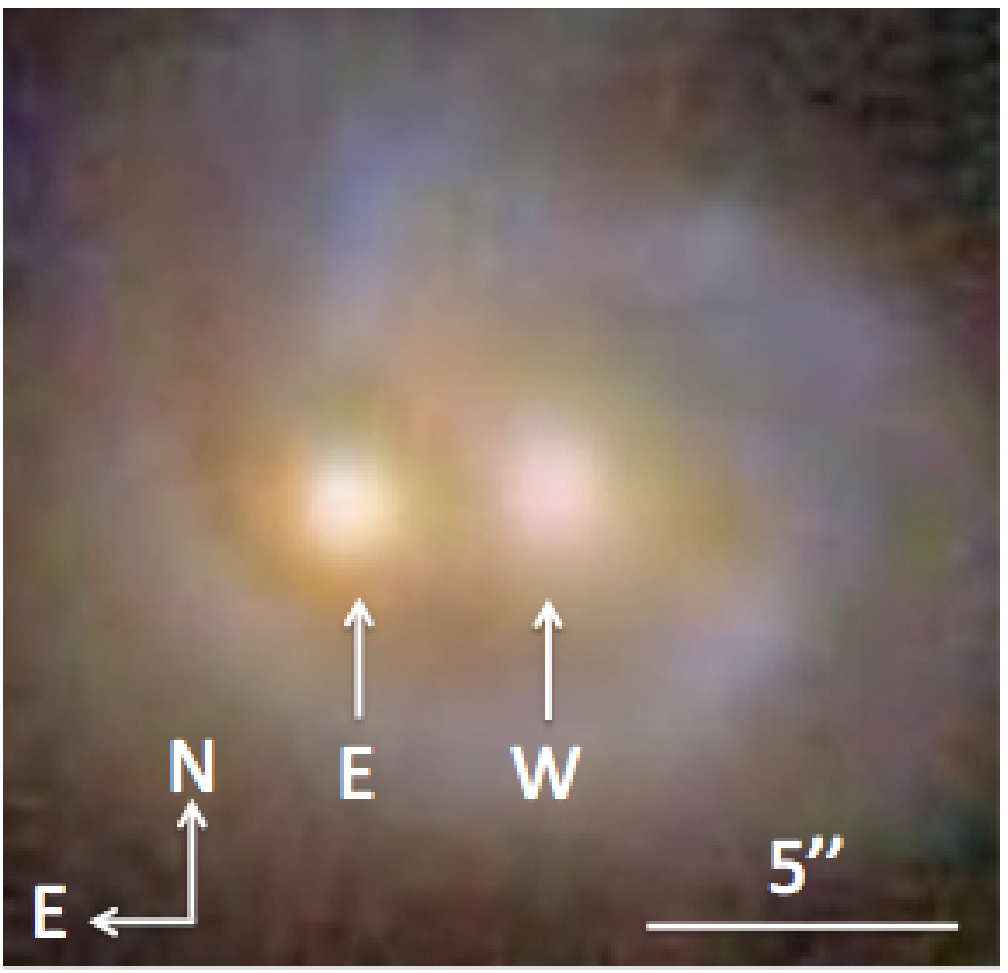} 
\includegraphics[width=7.7cm]{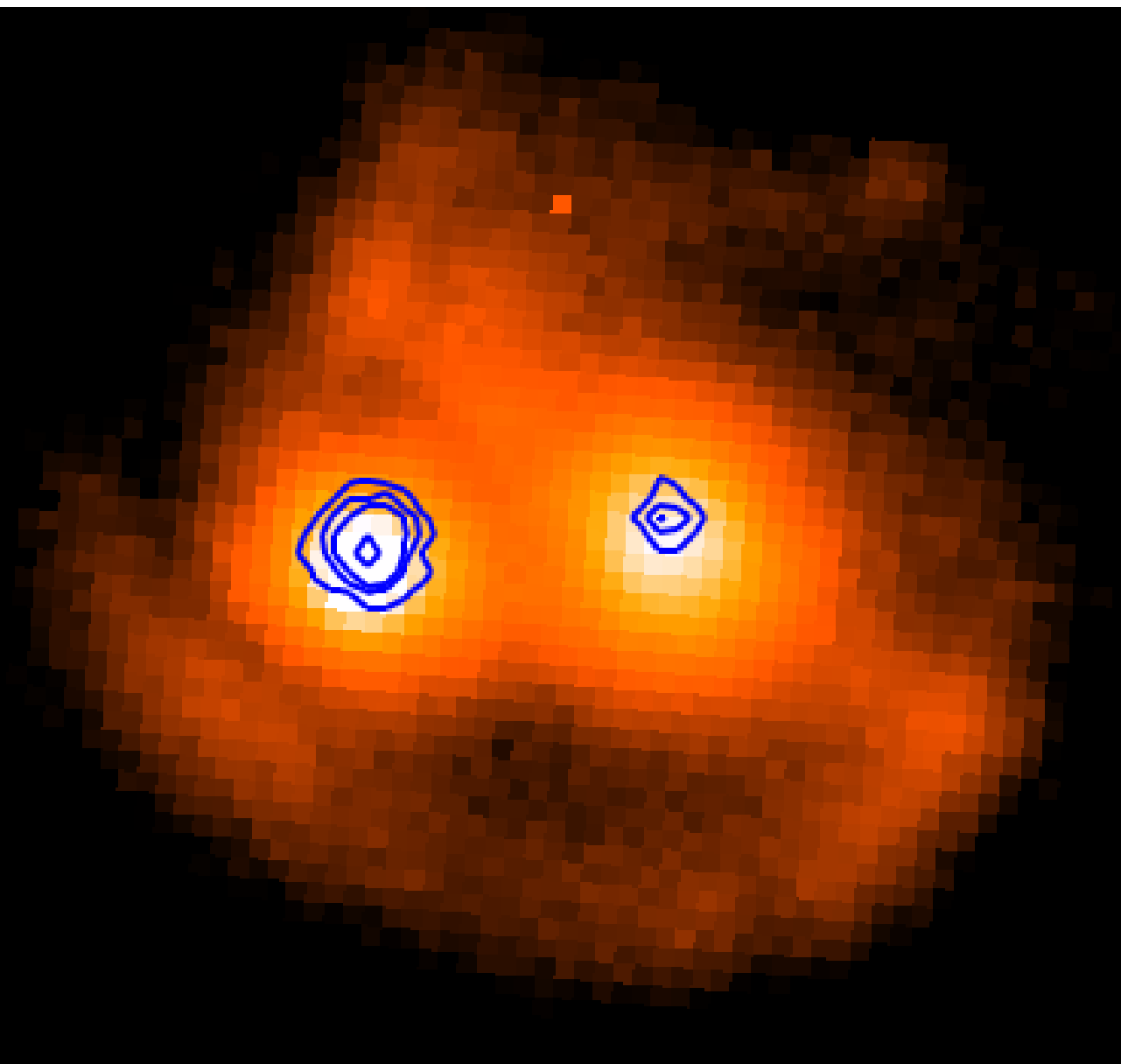} 
%\includegraphics[width=7.7cm]{tidaltail.jpg} 
%{\bf Upper right}: $\C$ hard band (2-8 keV) image of Mrk 739 showing two unresolved hard X-ray point sources. 
%{\bf Lower Right}:  Tidal tails provide important clues as to the stage and strength of the merger.  To look for low surface brightness tidal tails we have combined the griz images and smoothed by a factor of 3.  The two point sources to the NE are stars.  Based on the extended emission, we see low surface brightness tidal tails extending 30$\arcsec$ (17 kpc) to the NE and SW shown in the white contours.  In addition, we see a bar structure in Mrk 739W.
\caption{{\bf Top}:  composite SDSS \textit{gri} filter image of Mrk 739.   {\bf Bottom}:   SDSS $r$-band image overlaid with contours from $\C$ (blue) at same scale.  The X-ray sources are coincident with the bulge components seen in the optical for Mrk 739E and Mrk 739W.  }
\label{}
\end{figure}

	Two hard X-ray point sources coincide with the best fit model of the optical light from the bulge components (Figure~1).  The bulge magnitudes are $m_{r}$=14.03$\pm$0.15 for Mrk 739E and $m_{r}$=13.75$\pm$0.15 for Mrk 739W.  The small difference in apparent magnitudes suggests a major merger between the two galaxies.
	
\begin{figure*} 
\centering
\includegraphics[width=15cm]{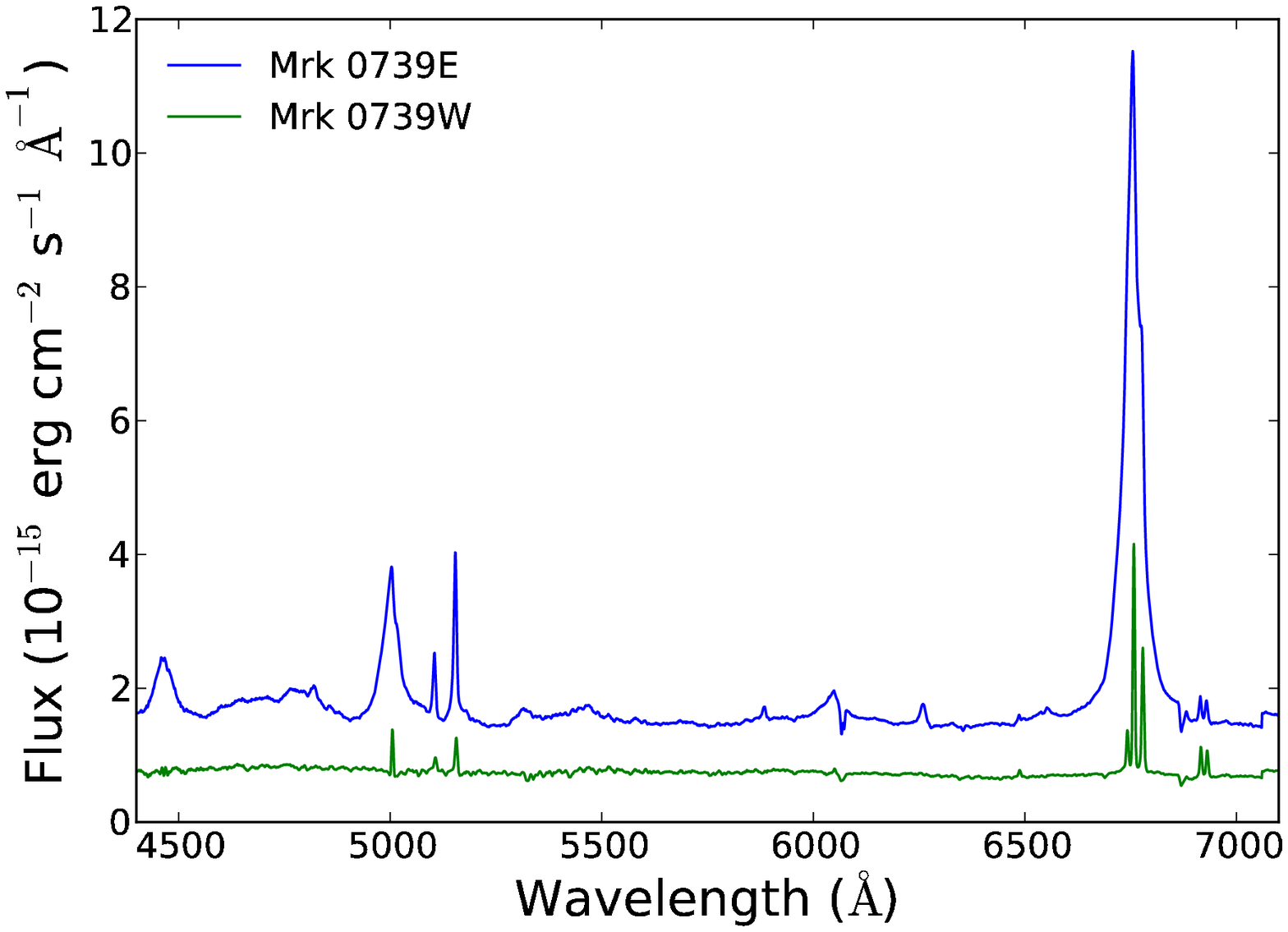} 
\includegraphics[width=5.4cm]{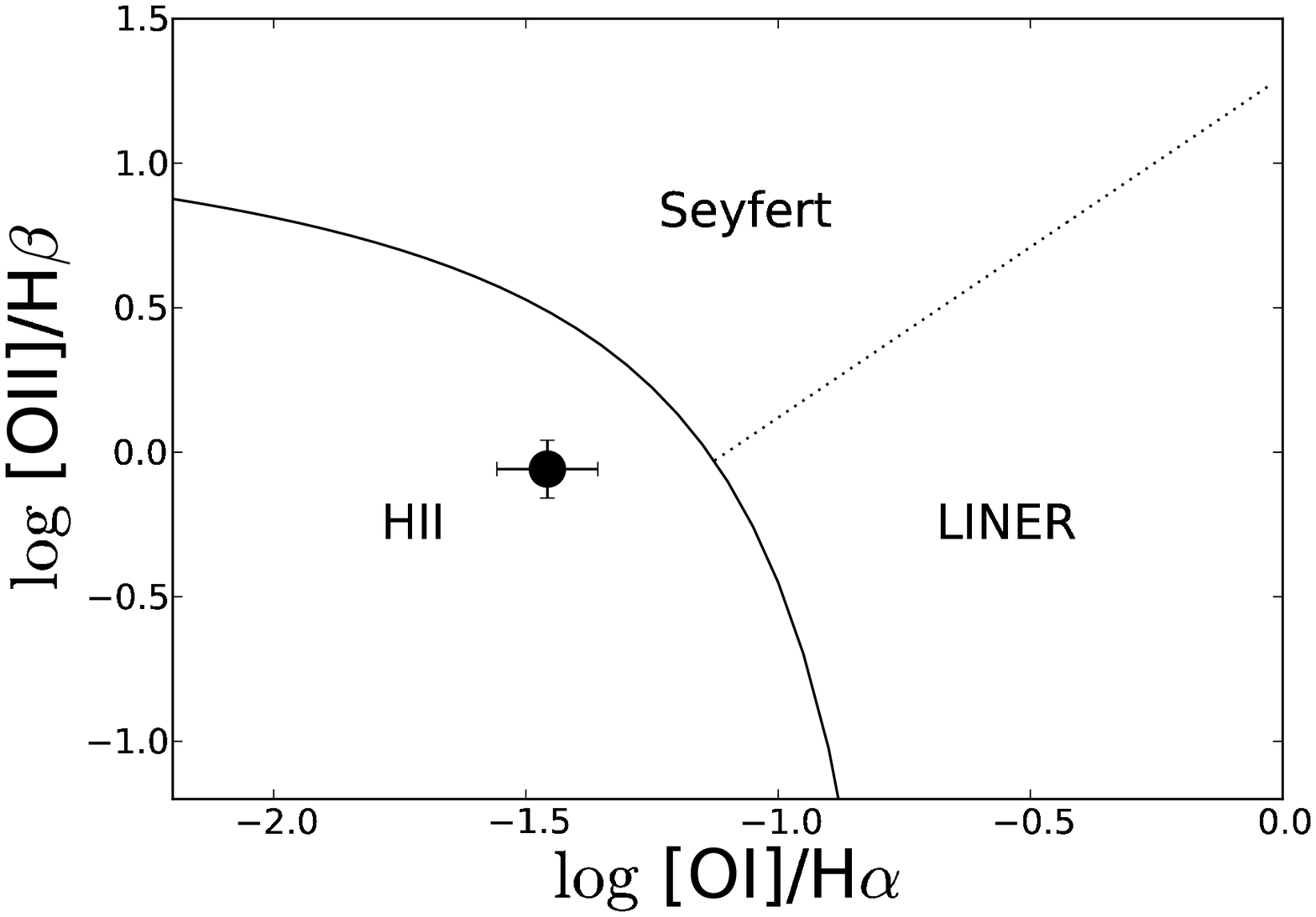} 
\includegraphics[width=5.4cm]{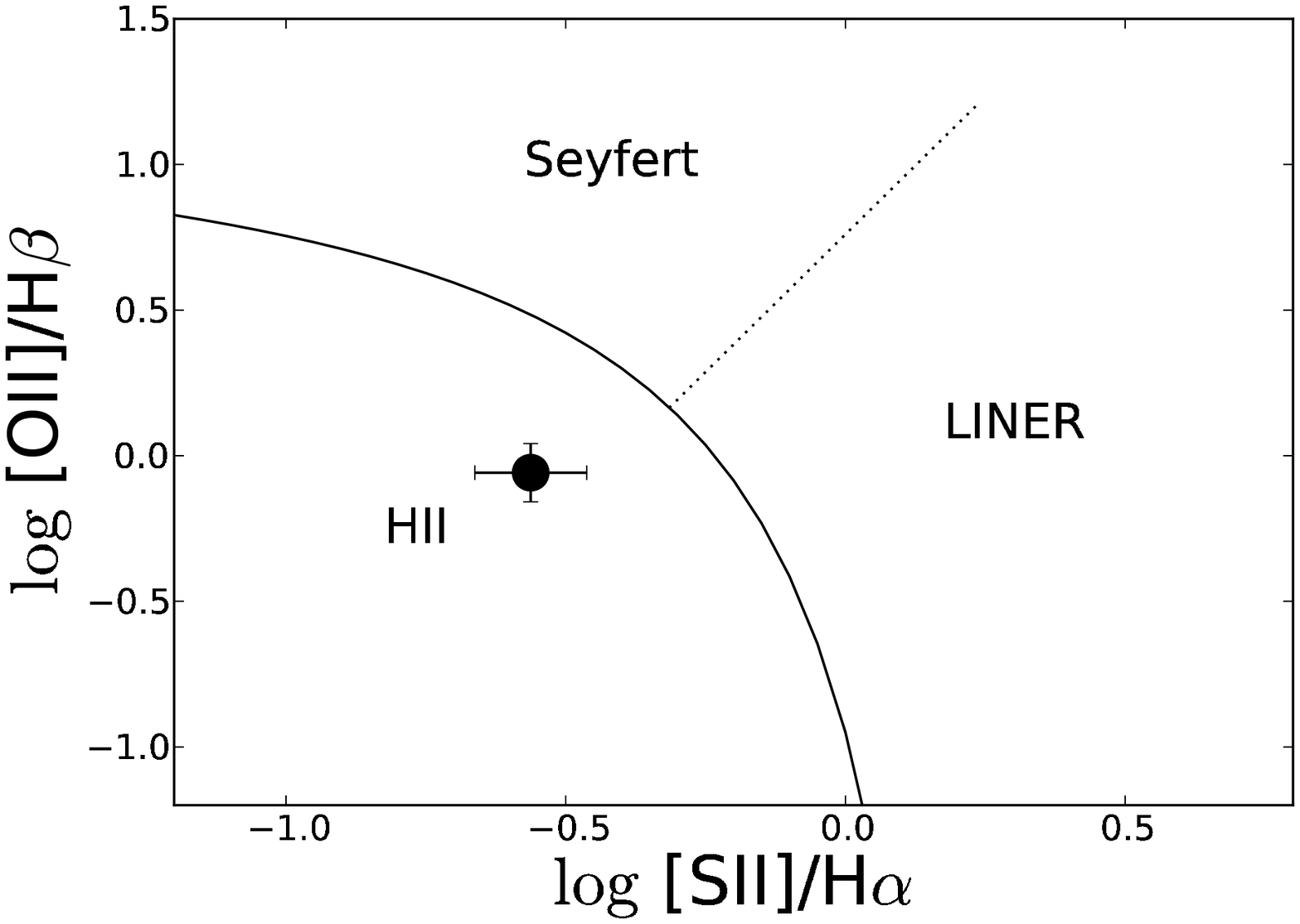} 
\includegraphics[width=5.4cm]{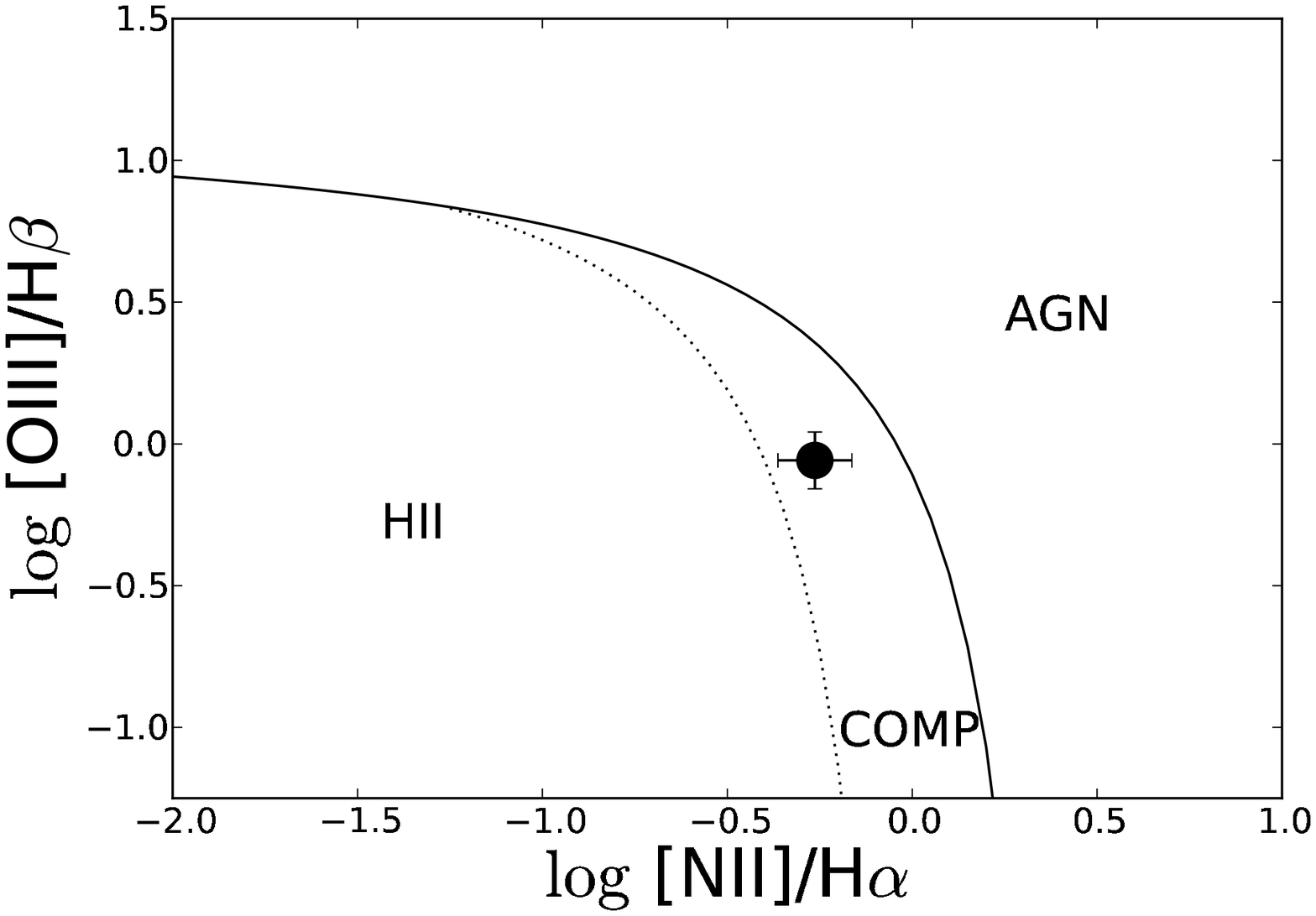} 
\caption{{\bf Upper panel}:  optical spectra of Mrk 739E and Mrk 739W.  {\bf Lower panel}:  emission line ratios of Mrk 739W, the narrow line source discovered to be an AGN in X-rays.  Mrk 739W is consistent with star formation in the [OI]/H$\alpha$ and [SII]/H$\alpha$ diagnostics and a composite galaxy in the [NII]/H$\alpha$ diagnostic.
}
\label{}
\end{figure*}

	The optical spectra for both sources are shown in Figure~2.  Mrk 739E shows broad lines (FWHM H$\beta$=2960 km s$^{-1}$ and H$\alpha$=2120 km s$^{-1}$) consistent with a  Seyfert 1.  Mrk 739E also has strong  [Fe VII] 5721 and 6087 $\AA$ emission,  a feature of some Seyfert 1 galaxies indicative of highly ionized material near the central AGN \citep[e.g.,][]{Veilleux:1988p12338}.  In Mrk 739W, there are narrow lines at the spectral resolution of instrument (FWHM=280 km s$^{-1}$).  Based on the Balmer decrement, $E(B-V)$=0.26 for Mrk 739E and $E(B-V)$=0.43 for Mrk 739W.  Mrk 739W is classified as a starburst using the [OI]/H$\alpha$ and [SII]/H$\alpha$ line diagnostics and a composite galaxy using the [NII]/H$\alpha$ diagnostic \citep{Kewley:2006p1554}.  Assuming that all of the H$\alpha$ luminosity is from star formation and using \citet{Kennicutt:1998p11796}, the estimated star formation rate (SFR)=0.3~$M_\odot$~yr$^{-1}$. 

%  Ho et al 2008 measure Hbeta of 42.25 and log Mbh of 7.11, Halpha FWHM of 1615 km s$^{-1}$.    
%This places it 4th among 53 broad line sources indicative of a very high Eddington ratio. 

\subsection{UV, Far-Infrared, and Radio}

\begin{figure} 
\centering
\includegraphics[width=7.7cm]{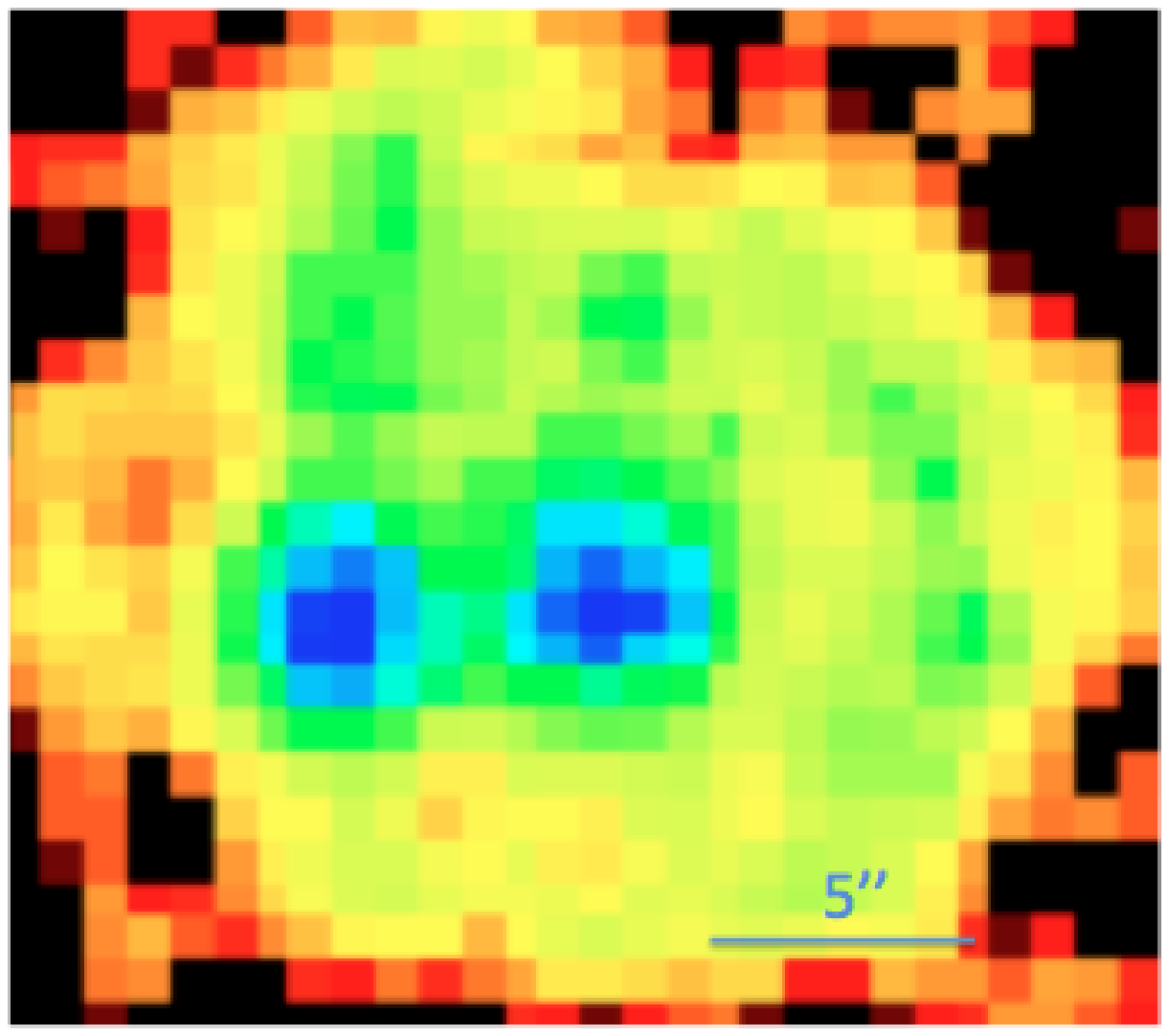}
\includegraphics[width=7.7cm]{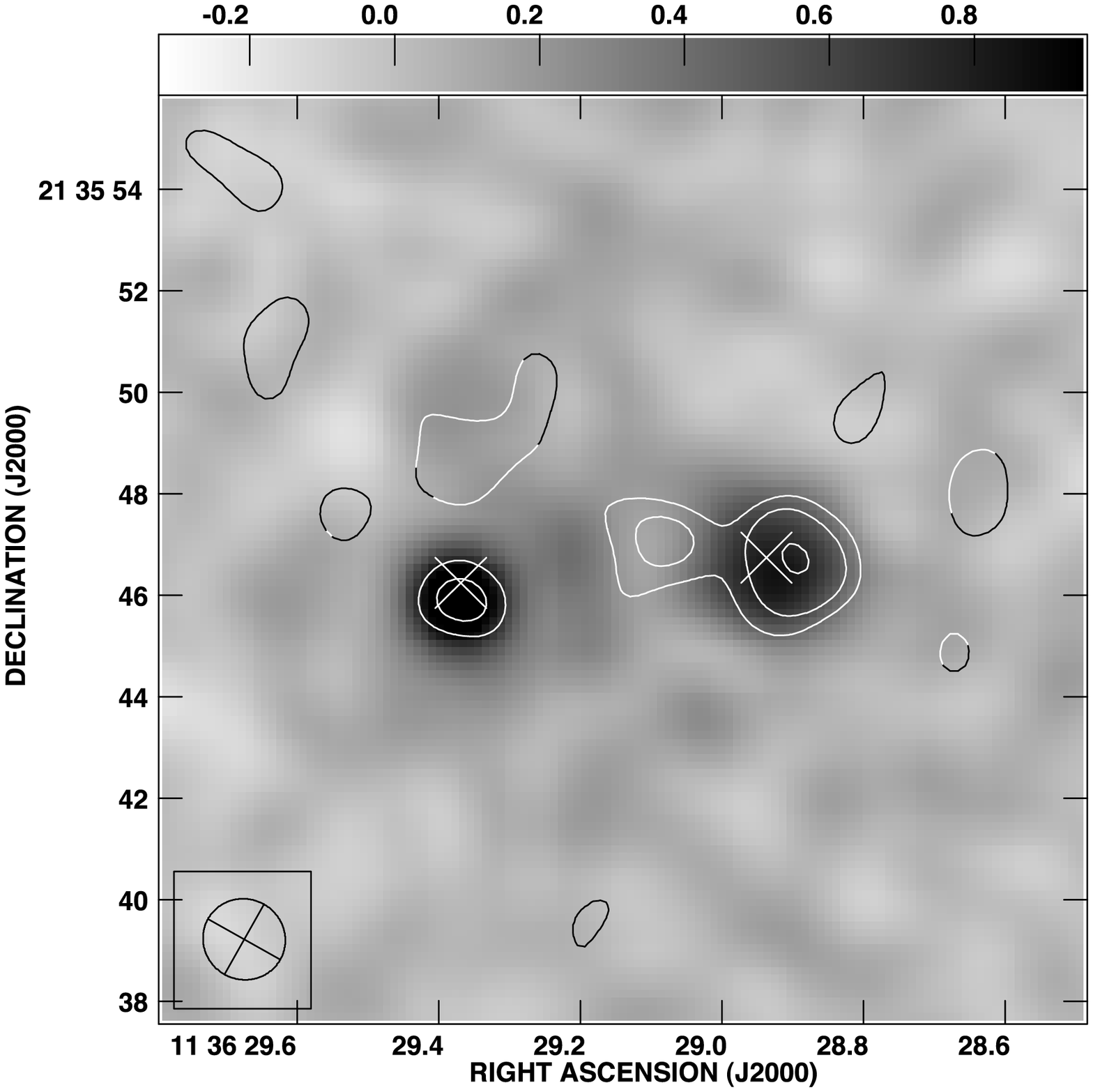} 
\caption{{\bf Top}:  $\X$ UVM2 image of Mrk 739. {\bf Bottom}:  the grey-scale image is the 1.49 GHz VLA data, while the contours are the 4.86 GHz data convolved to 1.49 GHz beam.  White $\times$'s indicate the $\C$ hard X-ray positions.  Both the UV and radio data in Mrk 739W show extended emission consistent with star formation.} 
\label{}
\end{figure}

	 The UV image from $\X$ Optical Monitor shows sources coincident with the hard X-ray sources (Figure~3).  In the UV,  $m_{\mathrm{UVW1}}$=15.4$\pm$0.1 and $m_{\mathrm{UVM2}}$=16.6$\pm$0.1 for Mrk 739E and  $m_{\mathrm{UVW1}}$=16.6$\pm$0.1 and $m_{\mathrm{UVM2}}$=16.4$\pm$0.1 for Mrk 739W.  The spectral index connecting 2500$\rm \AA$ and 2 keV, $\alpha_{\rm OX}$, is -1.10$\pm$0.04 for Mrk 739E, typical of an AGN \citep[-1.15$\pm$0.24;][]{Steffen:2006p11909}.  In Mrk 739W, $\alpha_{\rm OX}$=-1.53$\pm$0.1, suggesting the UV is dominated by star formation.  Assuming all the UV emission in Mrk 739W is from star formation and following \citet{Kennicutt:1998p11796}, we estimate SFR=0.6~$M_\odot$~yr$^{-1}$.  	 
	 
	Mrk 739 was detected in all bands by IRAS. The measured luminosity is $\log L_{\odot,\mathrm{IR}}$=10.9 and $\log L_{\odot,\mathrm{FIR}}$=10.6.  Following the Kennicutt (1998) relationship between far-infrared (FIR) luminosity and star formation rate, we estimate SFR=6.9~$M_\odot$~yr$^{-1}$.
	
	Mrk 739E was detected at both 1.49 and 4.86 GHz (Figure 3), with a flux density of 2.6$\pm$0.2 mJy at 1.49 GHz and 0.5$\pm$0.2 at 4.86 GHz (after convolving to match the 1.49 GHz resolution). This is a spectral index ($S_{\mathrm{\nu}} = K\times \nu^{-\alpha}$) of $\alpha$ =1.2$\pm$0.5.
	
	The VLA data for Mrk 739W are consistent with resolved star formation and show no signs of an AGN.  The emission at 1.49 GHz has an integrated flux density of 2.6$\pm$0.2 mJy.   The 4.86 GHz convolved data also shows resolved emission with an integrated flux density of 1.3$\pm$0.3 mJy.  The spectral index for Mrk 739W is $\alpha$ = 0.8$\pm$0.3, consistent with optically thin synchrotron emission from supernovae found in star forming galaxies.  We use \citet{Yun:2001p12041}  to convert the 1.49 GHz luminosity of Mrk 739W to a SFR of 3.1  $M_\odot$~yr$^{-1}$.
	
\subsection{X-ray}

	We detect two hard X-ray sources coincident with the eastern optical nucleus (Mrk~739E) and western nucleus (Mrk~739W).  Both sources show hard X-ray spectra extending out to 10 keV (Figure~4).  In the 2--10 keV band, we find a FWHM of 0.48$\arcsec\pm0.05$ (280 pc) for Mrk 739E and 0.51$\arcsec\pm0.07$ (295 pc) for Mrk 739W.  The $\C$ spectra of Mrk 739E is well fit ($\chi^2_{\nu}$=1.4) by a power law ($F\propto E^{-\Gamma+2})$ with photon index $\Gamma=2.1\pm0.1$  and $N_{\mathrm{H}}=1.5\pm0.2\times10^{21}$ cm$^{-2}$ consistent with a Seyfert 1.  Mrk 739W is well fit (C-stat/dof=0.8) by a harder power law with more absorption and a photon index $\Gamma=1.0\pm0.2$ and $N_{\mathrm{H}}=4.6\pm0.1\times10^{21}$ cm$^{-2}$.  While positive residuals do exist at the location of the neutral 6.4 keV iron K$\alpha$ line in Mrk 739W, there are too few counts to confirm its existence.  The 2--10 keV absorption-corrected luminosities are $L_{2-10 \: \mathrm{keV}}$ = 1.1$\times10^{43}$ and $1.0\times10^{42}$  erg s$^{-1}$ for Mrk 739E and Mrk 739W, respectively.  An archival $\X$ 2009 observation of Mrk 739 is unable to resolve the emission to either source, but shows $L_{2-10 \: \mathrm{keV}}$=1.0$\times10^{43}$ and  $\Gamma=1.92\pm0.02$ consistent with the $\C$ spectra of Mrk 0739E.
	
\begin{figure*} 
\centering
\includegraphics[width=8.17cm]{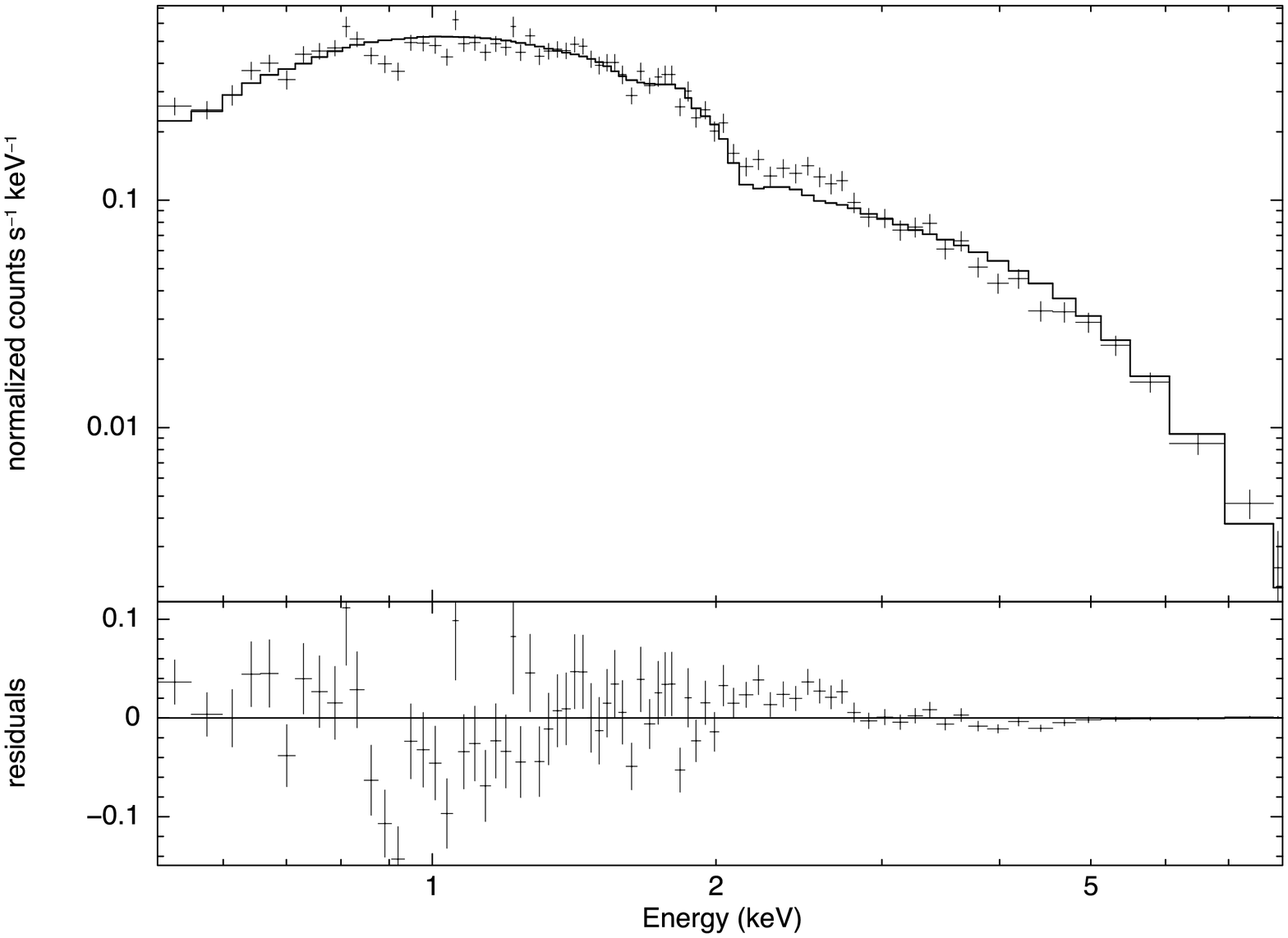} 
\includegraphics[width=8.17cm]{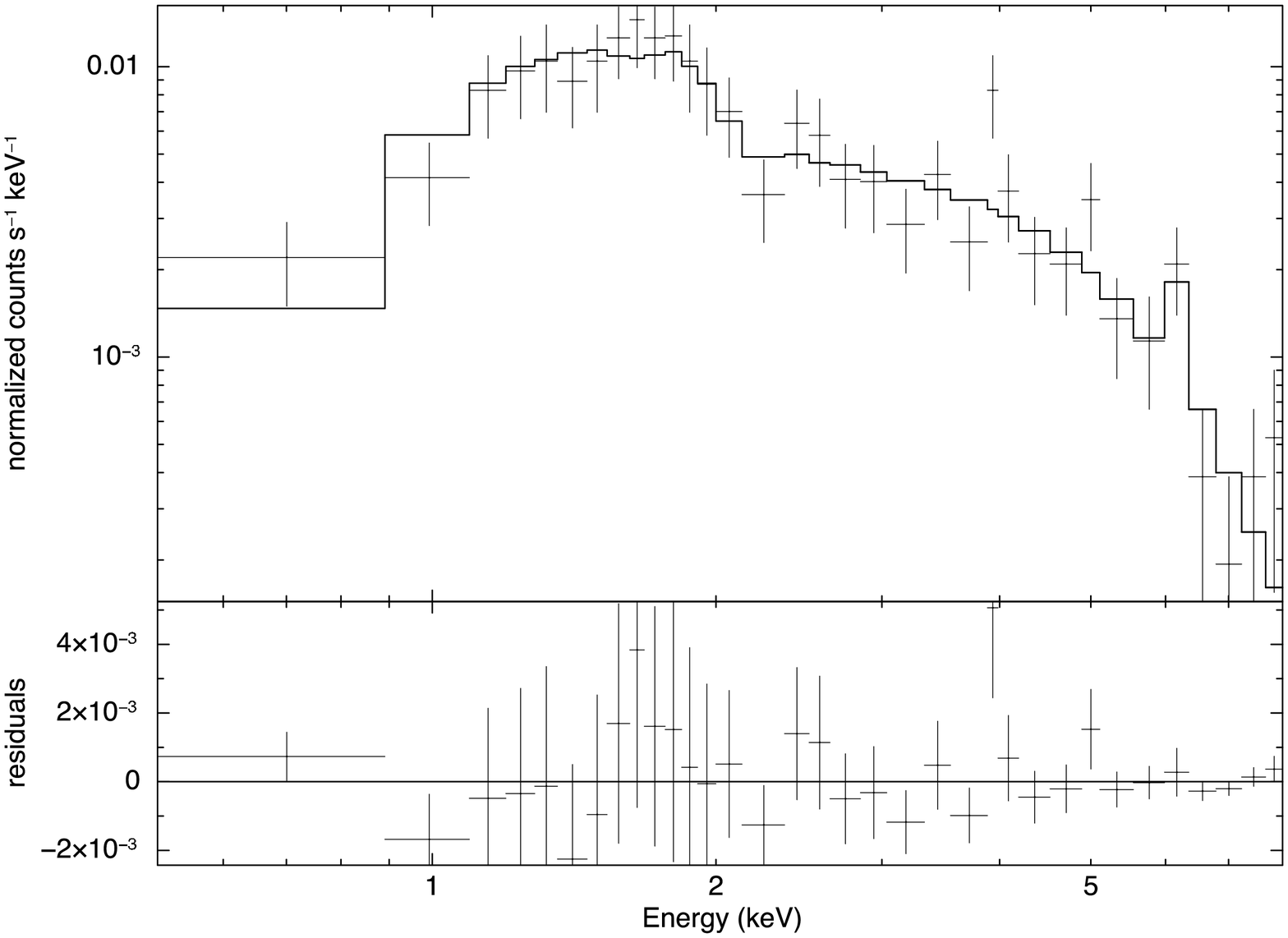} 
\includegraphics[width=8.1 cm]{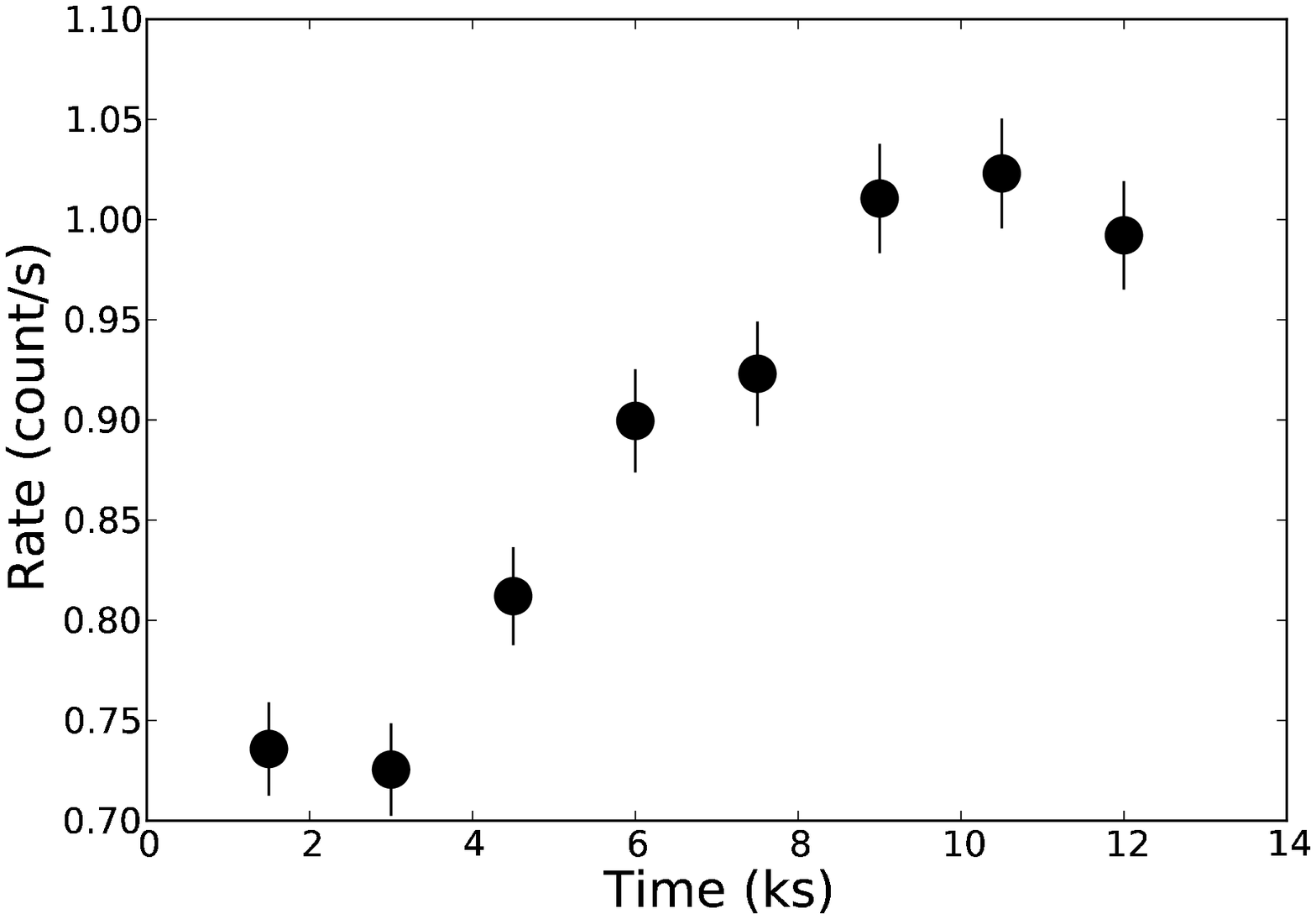}
\includegraphics[width=8.1 cm]{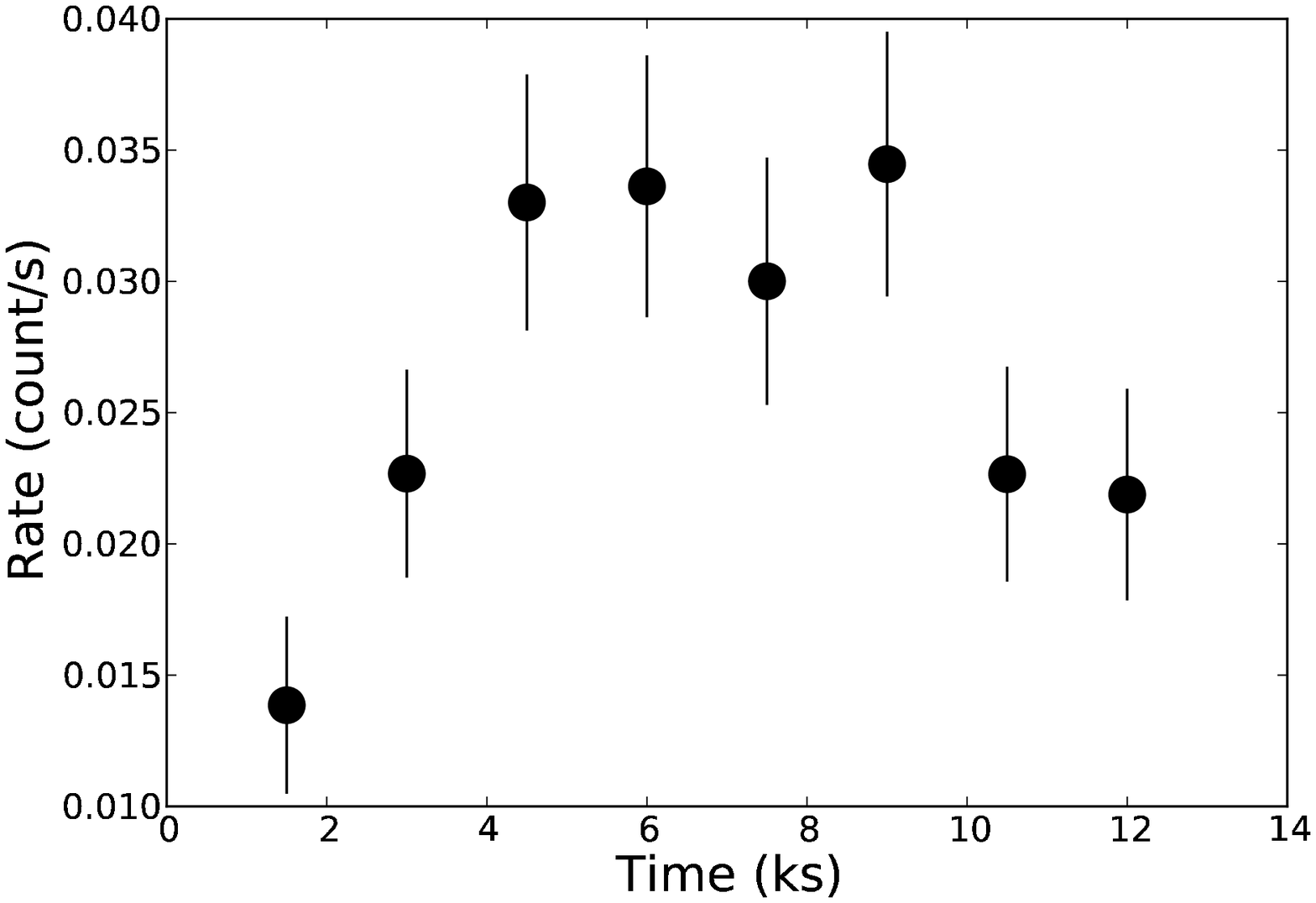} 
\includegraphics[width=8.1 cm]{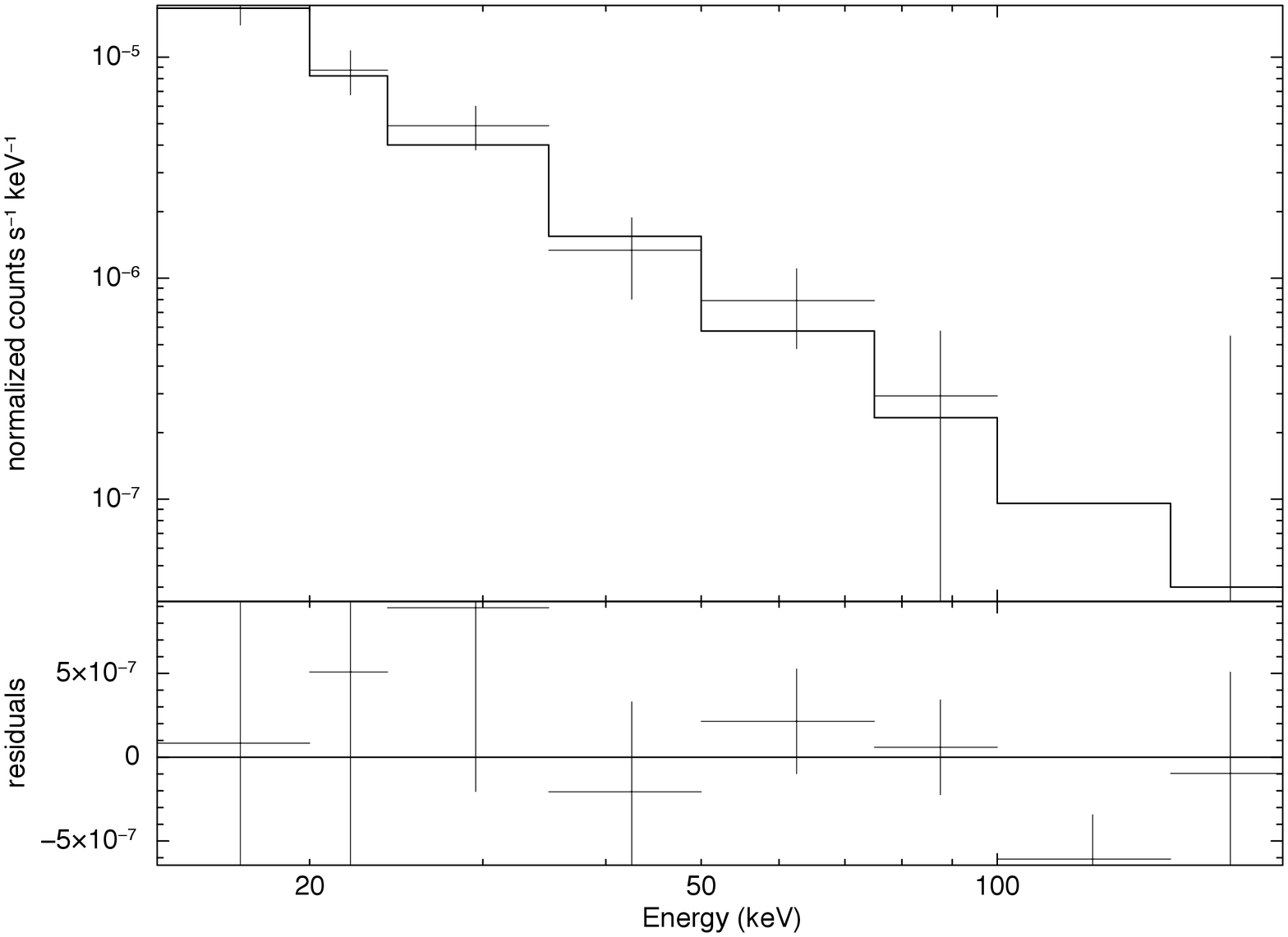} 
\includegraphics[width=8.1 cm]{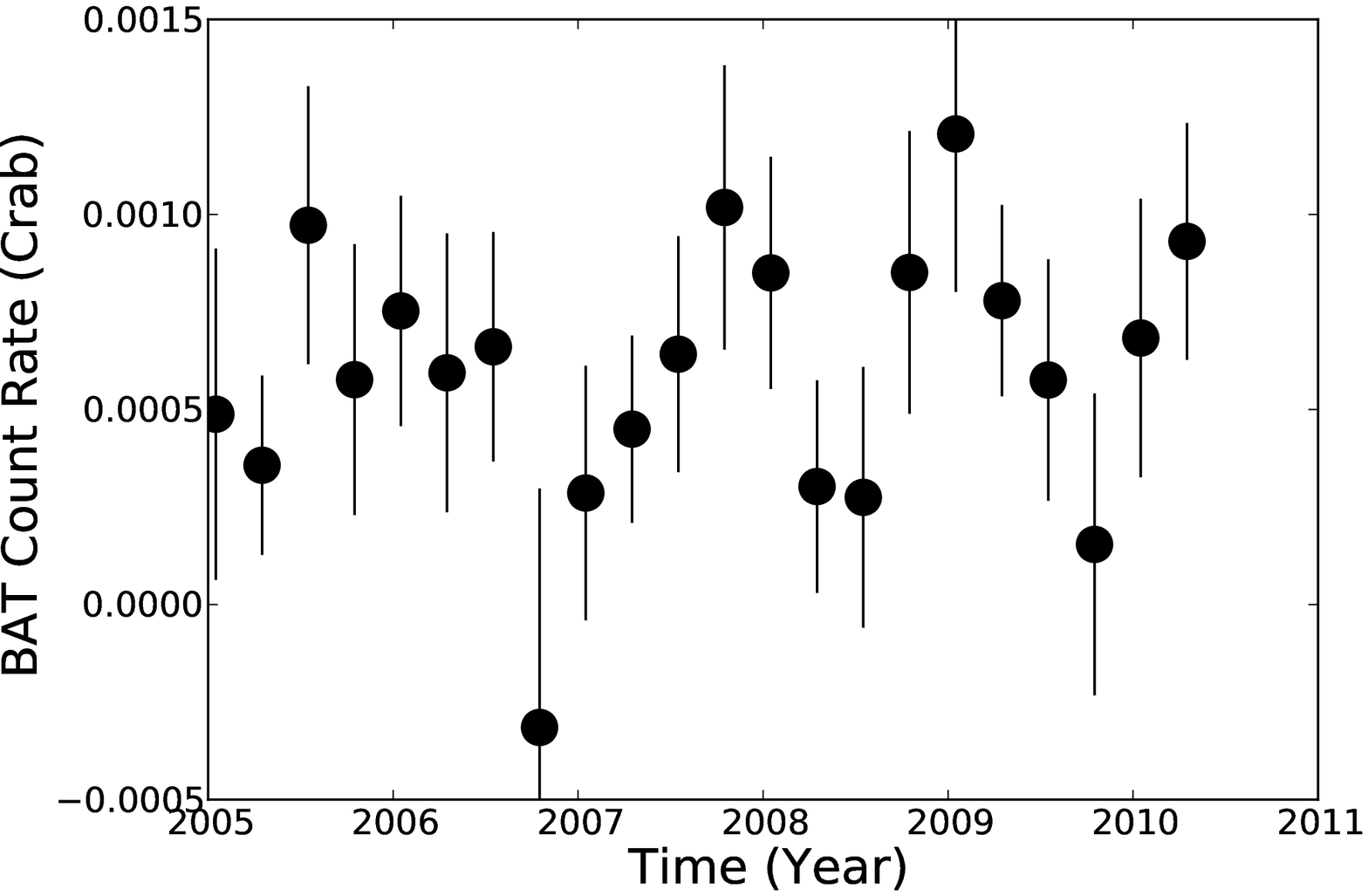} 

\caption{{\bf Upper panel}: $\C$ spectra of Mrk 739E (left) Mrk 739W (right). {\bf Middle panel}:  $\C$ count rate in 1.5 ks bins for Mrk 739E (left) Mrk 739W (right).  There is a factor of 2.5 variability in flux over the 4-hour observation of Mrk 739W.  {\bf Lower panel}: $Swift$ BAT spectra (left) and average count rate in 3 month bins (right) for Mrk 739. }
\label{}
\end{figure*}
	
	Timing analysis is a critical part of identifying AGN since nearly all show variability.  Bins of 1.5 ks were chosen to ensure $>$20 counts per bin (Figure~4).  We find statistically significant variations in the fluxes of both sources.   There is a factor of 2.5 change in flux for Mrk 739W and 0.3 for Mrk 739E during the $\approx$4-hour observation.
	
	Mrk 739 was detected by $Swift$ BAT with  $L_{14-195 \: \mathrm{keV}}$=2.4$\pm0.5\times10^{43}$ erg s$^{-1}$, $\Gamma$$=$$2.6\pm0.4$, and a factor of two variability over five years (Figure~4).  Because of the variability and much steeper photon index of Mrk 739W, it is difficult to identify the source of the $L_{14-195 \: \mathrm{keV}}$ luminosity (Mrk 739E, Mrk 739W, or both).

	%The resolution of $Swift$ BAT (120$\arcsec$) is unable to constrain the emission to either source.   

	%Mrk 739 was detected by $Swift$ BAT with $L_{14-195 \: \mathrm{keV}}$=2.4$\pm0.5\times10^{43}$ erg s$^{-1}$.  The resolution of $Swift$ BAT (120$\arcsec$) is unable to constrain the emission to either source.   Assuming the photon index and 2--10 keV luminosity of Mrk 739E, we estimate a  $L_{14-195 \: \mathrm{keV}}=1.2\times10^{43}$ erg s$^{-1}$.  Assuming the photon index and 2--10 keV luminosity of Mrk 739W, we estimate a  $L_{14-195 \: \mathrm{keV}}=2.2\times10^{43}$ erg s$^{-1}$.  Since the luminosity of Mrk 739E is insufficient to account for all of the measure ultra hard X-ray luminosity suggest Mrk 739W may contribute a significant part of the emission as a Compton-thick source.

%The absorption-corrected luminosities in the 0.5--10 keV band are 1.7$\times10^{43}$ and $1.0\times10^{42}$  erg s$^{-1}$ for the western and the eastern nucleus, respectively.

\subsection{Detection of a Binary AGN}

	The hard X-ray ($>$2 keV) band provides one of the best tools for finding AGN since it is less affected by contamination and absorption and can only be produced in large amounts by AGN.  Our discussion of this binary AGN will be limited to Mrk 739W since Mrk 739E was already known to be an AGN based on its BLR \citep{Netzer:1987p11560}.  
	
	%We also discuss why Mrk 739W may not be detected as an AGN at other wavelengths.  
	
	In Mrk 739W, the hard X-ray emission is point-like at the location of one of the bulge components in the galaxy consistent with an AGN.  Energetic phenomena related to vigorous star formation such as OB stars, X-ray binaries, and SN shocks produce hard X-rays.  However, in star forming regions the dominant X-ray emission is from point-like ultraluminous X-ray sources \citep[ULXs;][]{Bertram:2007p10319}.  ULXs are by definition not located at the centers of galaxies where the central supermassive black hole resides.
  
	 The luminous hard X-ray emission, hard photon index, and time variability of Mrk 739W also provide little support for the hypothesis that this source is a ULX.  In a study of 154 ULXs, the average luminosity is much lower ($L_{0.5-8 \: \mathrm{keV}}=  2.2\pm4.5\times10^{39} $ erg s$^{-1}$).  The most luminous ULX ever detected,  ESO 243-49 HLX-1 \citep{Farrell:2009p11689}, has a 0.2--10 keV luminosity of 10$^{42}$ erg s$^{-1}$, however, its hard X-ray (2--10 keV) luminosity is only 4$\times10^{40}$~erg~s$^{-1}$ because it has a soft photon index of 3.4.  Therefore, if Mrk 739W is a ULX, it is the most luminous ULX in the hard X-rays ($L_{2-10 \: \mathrm{keV}}$) by over an order of magnitude.  The average ULX power law index is also $\Gamma$=1.97$\pm$0.11, which is significantly softer than Mrk 739W ($\Gamma$=$1.0\pm0.2$).  Finally, no ULXs have shown such high amplitude variability over the short timescale of hours with this level of variability only seen on time scales of days to weeks \citep{Gladstone:2010p11983}. 

%86$\%$ of these ULXs showed no indication of time variation in luminosity whereas Mrk 739W shows a factor of 2.5 variability over 4-hours.
	 
%We find that the Balmer decrement corrected H$\alpha$ is $L_{H\alpha}=3.8\times10^{40}$.  

	The measured SFRs in Mrk 739W provide an additional constraint as to whether the hard X-ray emission could be from star formation.  A SFR greater than 200 $M_\odot$~yr$^{-1}$ would be required to generate the observed hard X-ray luminosity based on the relationship between SFR and X-ray emission \citep{Ranalli:2003p11744}.  The predicted SFR in Mrk 739W is 0.3, 0.6, and 3.1 $M_\odot$~yr$^{-1}$ from the H$\alpha$, UV luminosity, and 1.4 GHz emission.  The predicted SFR from FIR emission of the combined system is 6.9 $M_\odot$~yr$^{-1}$.  All of these rates are significantly lower than the 200 $M_\odot$~yr$^{-1}$ needed to generate the observed hard X-ray luminosity.  In addition, it is likely that much of this star formation would be extended and resolved in $\C$.
	
	It is interesting that Mrk 739W has not been detected as an AGN using optical emission line spectroscopy.  \citet{Noguchi:2010p7451} found that optical emission line studies are biased against 'buried AGN' that have a small scattering fraction or a small amount of narrow line region gas.  AGN with a low ratio of [O III] to hard X-ray luminosity ($L_{[\mathrm{O III}]}$/$L_{2-10 \: \mathrm{keV}}<$0.1) tend to be 'buried AGN'.  The $L_{[\mathrm{O III}]}$/$L_{2-10 \: \mathrm{keV}}$=0.008, consistent with a 'buried AGN' and the lowest ratios found in their study.  This finding is also consistent with a recent study that found that merging AGN selected in the ultra hard X-rays tend to have low $L_{[\mathrm{O III}]}$/$L_{14-195 \: \mathrm{keV}}$ ratios and be preferentially misclassified using optical line diagnostics \citep{Koss:2010p7366}.	
	
	For AGN with low luminosity in the [O III] line, nebular emission from star formation can overwhelm the AGN signature in optical emission line diagnostics.   \citet{Schawinski:2010p6049} found that for $L_{[\mathrm{O III}]}$=10$^{40}$ erg s$^{-1}$, nearly 54\% of star forming galaxies with AGN will be classified as star forming or composites.  The small value of $L_{[\mathrm{O III}]}$=$7.5\times10^{39}$ erg s$^{-1}$ in Mrk 739W suggests that star formation is overwhelming the AGN photoionization signature.

	\subsection{CO Properties and Molecular Gas Mass}

	CO velocity profiles can provide information on the dynamics of the molecular gas.  The $^{12}$CO 3--2 and 2--1 spectra in Mrk 739 have almost identical shapes.  Each spectrum has a narrow profile with FWHM=94$\pm$8  and 98$\pm$6 km s$^{-1}$ for single-Gaussian fits to the CO (2--1) and CO (3--2) profiles, respectively. These profiles are significantly narrower than the CO (2--1) emission from NGC 6240 (Figure 5) and imply a nearly face-on orientation to any disk-like structure in this system.
	
	CO measurements also provide estimates of the amount of molecular gas.  In Mrk 739, $I_{\mathrm{CO}}$=109$\pm$33 and 169$\pm$51 Jy km s$^{-1}$ for the 2--1 and 3--2  lines.  Following \citet{Solomon:1992p11420}, $L'_{\mathrm{CO}}$=10.9$\times10^{8}$ and 7.5$\times10^{8}$ K km s$^{-1}$ pc$^{-2}$ for the 2--1 and 3--2 lines.  Adopting $\alpha$=1.5-4 $M_\odot$ (K km pc$^{-2}$)$^{-1}$ for the conversion from CO luminosity to molecular hydrogen, we find $\log$~$M_\odot$(H$_2$)=9.2-9.6, similar to the Milky Way \citep{Sanders:1984p12189}.

\subsection{Relative Velocity of the Binary AGN}

	Measurements of radial velocities provide important insights about the dynamics of the merger.  We use the Na I $\lambda\lambda$ 5890, 5896 (Na D) absorption lines from stars and cold gas since narrow emission lines in AGN often have blueshifts \citep{Bertram:2007p10319}.    There is an offset of $\approx$40 km s$^{-1}$ between the two bulge components (8995$\pm$15 km s$^{-1}$ for Mrk 739E and 8953$\pm$15 km s$^{-1}$ for Mrk 739W).  The CO data also show evidence of two components with the peak brightness temperatures similar to the radial velocities in the Na D absorption lines (Figure 5).  When fit with gaussians, the peaks consistent with the Na D radial velocities (8921$\pm$22 and 8980$\pm$16 km s$^{-1}$ in CO 2--1 and 8956$\pm$12 and 8993$\pm$22 km s$^{-1}$).  High resolution (1$\arcsec$) interferometric CO imaging of this system would provide evidence to confirm this picture.
	
	There is also evidence of outflows in the narrow line region of Mrk 739E.  There is a 192$\pm22$ km s$^{-1}$ blueshift in the [O III] line and a 153$\pm$25 km s$^{-1}$ in the lower ionization [O I] $\lambda$ 6300 line compared to the Na D absorption.  This blueshift is consistent with other nearby QSOs which have an average [O III] blueshift of -174 km s$^{-1}$ \citep{Bertram:2007p10319}.     In Mrk 739W, there is no evidence of outflows in the narrow line region.

\subsection{Bolometric Luminosity and Eddington Ratios}
	Using $\C$ and UV photometry and following \citet{Vasudevan:2009p7223}, the bolometric luminosity is 1.0$\times$10$^{45}$ erg s$^{-1}$ in Mrk 0739E (Figure 5).  The extinction corrected 2500 $\AA$ luminosity is $\log L_{\mathrm{2500\AA}}=43.7\pm0.3$.  Using H$\beta$ and continuum emission \citep{Vestergaard:2006p11438}, the black hole mass is log $M_{\mathrm{BH}}$=7.04$\pm$0.4,  giving an Eddington ratio of $\lambda_{\mathrm{Edd}}$=0.71, the highest amongs all the $Swift$ BAT selected AGN \citep{Vasudevan:2010p5970}.   Our estimates are consistent with \citet{Ho:2008p11625} who find $\lambda_{\mathrm{Edd}}$=0.78 using only optical spectra and the same method to determine black hole mass.    Uncertainties in intrinsic dust reddening, as well as the inclination angle and spectral hardening parameter in the accretion disk model can lower the Eddington ratio at most 58\% to $\lambda_{\mathrm{Edd}}$=0.30.

	In Mrk 739W,  the bolometric luminosity is 2$\times$10$^{43}$  erg s$^{-1}$ using only the hard X-ray data with a bolometric correction factor of 22 from \citet{Vasudevan:2009p7223}.

\begin{figure*} 
\centering
\includegraphics[width=7.7cm]{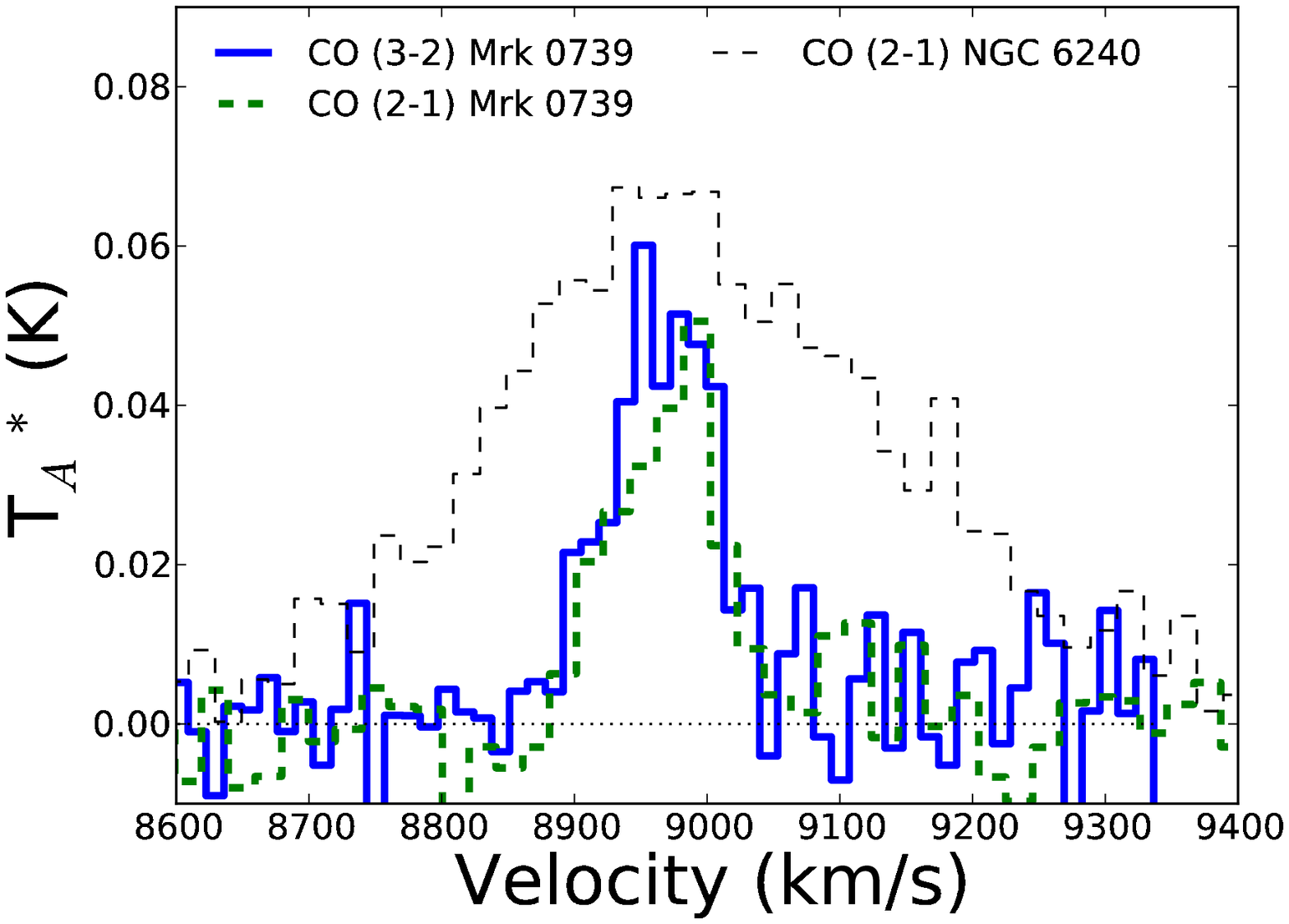} 
\includegraphics[width=7.7cm]{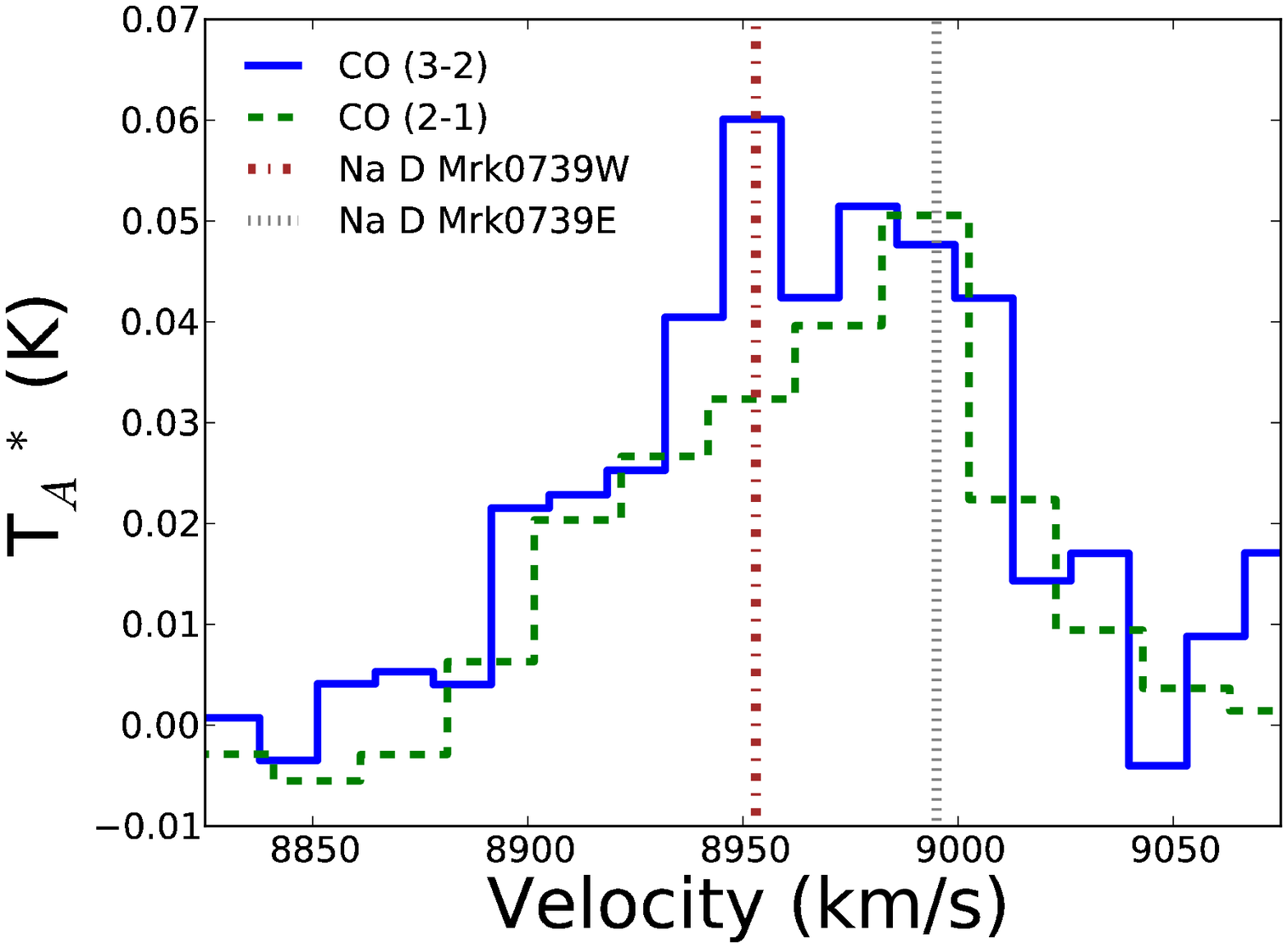} 
\includegraphics[width=7.7cm]{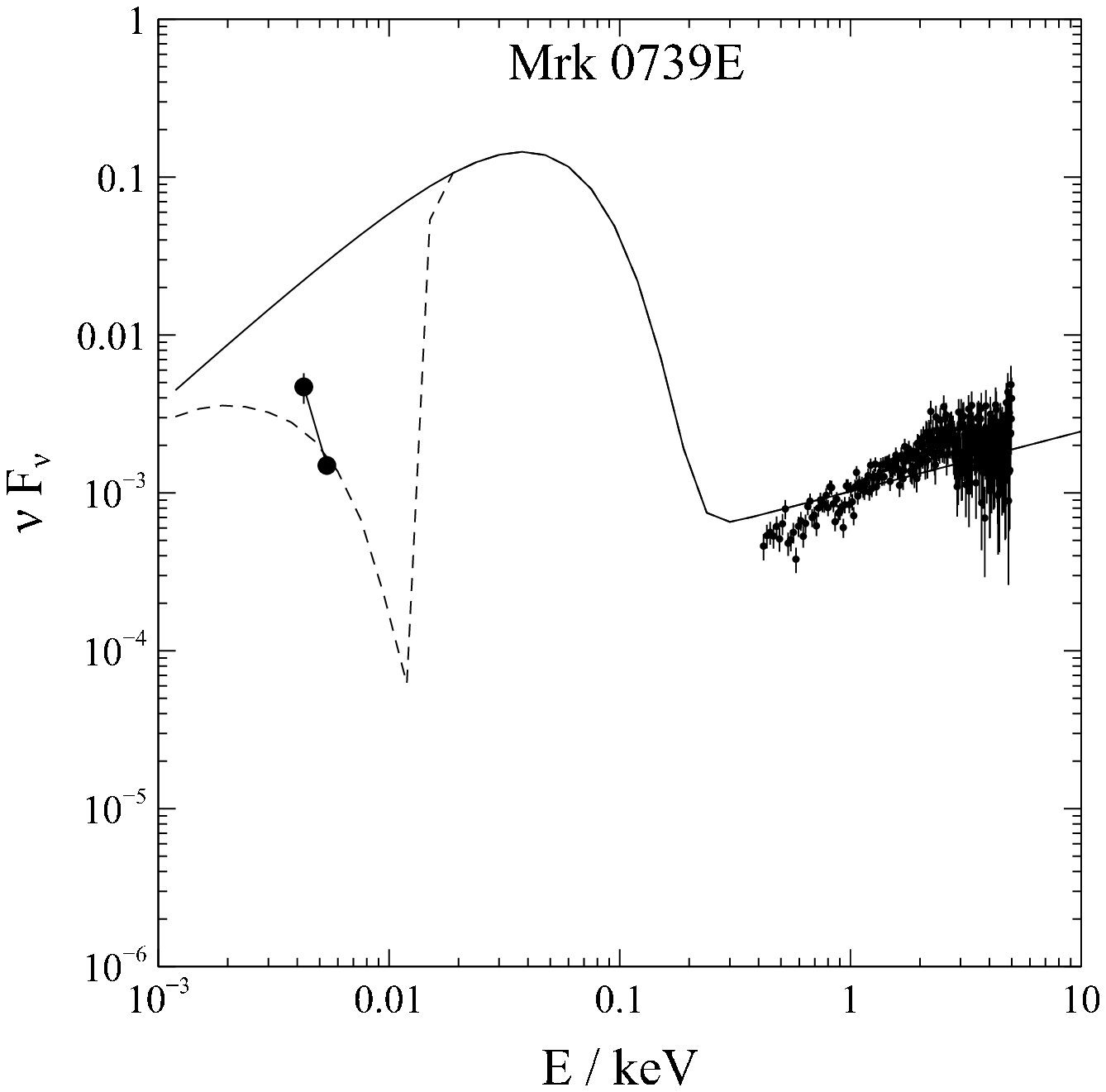}
\caption{{\bf Upper left}: CO observation of Mrk 739 with an observation of NGC 6240 overlaid for comparison.  The  brightness temperature of NGC 6240 is reduced by 0.63 to account for the increased distance of Mrk 739  \citep{Solomon:1992p11420}.  {\bf Upper right}:  CO observation with radial velocities of the Na D absorption lines from the optical spectra. {\bf Lower}:  measured UV and X-ray emission for Mrk 739E.  The extinction-corrected SED model is shown as a solid line and the dashed line indicates the fit to the observed data.}
\label{}
\end{figure*}

	%The extinction corrected $\alpha_{\rm OX}$=-1.53$\pm$0.26 compared to a value of -1.28$\pm$0.24 expected for an AGN of this luminosity \citep{Steffen:2006p11909}.	

%To avoid double counting, we assume that any longer-wavelength emission in the IR is due to dust reprocessing and is not included in the power output. 
%Mrk 739E also has an available SDSS spectrum with coverage of the infrared calcium triplet.   ( 8770$\pm$10 km s$^{-1}$) for Mrk 739E and 5155.7$\pm0.2\mathrm{\AA}$ (8910$\pm$10 km s$^{-1}$) for Mrk 739W. 

\section{Discussion}	  
	
	We discovered a binary AGN in the galaxy Mrk 739 based on $\C$ imaging showing two unresolved (FWHM$\approx$300 pc) luminous hard X-ray sources with a projected separation of 3.4 kpc (5.8$\pm$0.1$\arcsec$).   We find that a high level of star formation combined with a very low $L_{\mathrm{[O III]}}$/$L_{2-10 \: \mathrm{keV}}$ ratio cause the AGN to be missed in optical spectroscopy.  In the radio, there is resolved emission with a spectral index consistent with star formation.  The CO observations of the (3--2) and (2--1) lines indicate large amounts of molecular gas in the system. This gas could be driven towards the black holes during the violent galaxy collision and be key to fueling the binary AGN.  In Mrk 739E, there is a high Eddington ratio ($\lambda_{\mathrm{Edd}}$=0.71) and small black hole ($\log \mathrm{M}_{\mathrm{BH}}=7.05\pm$0.3) consistent with an AGN accreting at a high accretion rate.  Other than NGC 6240, this stands as the clearest and nearest case of a binary AGN discovered to date. 

	Mrk 739 is an important example of how critical high resolution ($<1\arcsec$), hard X-ray ($>$2 keV) imaging is in finding binary AGN ($<5$ kpc).  Observations with $\C$ provide one of the most effective tools since obscuration and/or contamination from merger induced star formation can hide the AGN at other wavelengths.  Mrk 739W showed no evidence for hosting an AGN until the $\C$ observation despite a host of previous observations including UV and optical spectroscopy, and radio data from the VLA.   While mega surveys such as the SDSS,  2dF, and 6dF are finding valuable information on hundreds of thousands of AGN using optical emission line diagnostics, this technique can be biased against finding AGN in objects that have high levels of star formation or obscuration like mergers   \citep[see also][]{Koss:2010p7366,Goulding:2009p6170,Veilleux:2009p9544}.
  
	The three nearest binary AGN (NGC 6240, Mrk 739, Mrk 463) discovered to date with $\C$ share many properties which may hold clues as to why they form.  A surprising result is that all three are luminous in the ultra hard X-rays and detected in the $Swift$ BAT all sky survey  (log $L_{14-195 \: \mathrm{keV}}>43.4$ erg s$^{-1}$).  This suggests that binary AGN happen more often in systems with bright X-ray AGN.  More X-ray follow-up work needs to be done with less luminous merging systems to confirm this result.  The large FIR (60 and 100 $\micron$) luminosities  (log $L_{\odot,\mathrm{FIR} }$=11.5, 11.1, and 10.6 for NGC 6240, Mrk 463, and Mrk 739) suggest that these systems may be linked to gas-rich progenitor galaxies consistent with theoretical models (Yu et al.~2011, submitted).

{\it Facilities:}  \facility{CXO}, \facility{Swift}, \facility{Sloan}, \facility{IRAS}, \facility{Gemini:GMOS}, \facility{JCMT}, \facility{VLA}
{\it Facilities:}  \facility{Swift}, \facility{Sloan}, \facility{KPNO:2.1m}, \facility{IRAS}, \facility{AKARI}
\bibliographystyle{/Applications/astronat/apj/apj}
\bibliography{/Applications/astronat/bibfinal}
\expandafter\ifx\csname natexlab\endcsname\relax\def\natexlab#1{#1}\fi

\end{document}